\documentclass[12pt]{article}
\usepackage[english]{babel}

\usepackage{amssymb,amsmath}

\usepackage{amsfonts}
\usepackage{latexsym}
\usepackage{graphicx,color}
\usepackage{epsfig}
\usepackage{verbatim}
\usepackage{amsthm}
\usepackage{amsbsy}

\usepackage{hyperref}
\usepackage{cite}
\definecolor{link}{rgb}{.8,.15,.1}
\hypersetup{colorlinks=true,linkcolor=link,citecolor=link,urlcolor=link,linktocpage}
\newcommand{\be}{\begin{equation}}
\newcommand{\ee}{\end{equation}}
\newcommand{\bi}{\begin{itemize}}
\newcommand{\ei}{\end{itemize}}
\newcommand{\bea}{\begin{eqnarray}}
\newcommand{\eea}{\end{eqnarray}}
\newcommand{\ba}{\begin{array}}
\newcommand{\ea}{\end{array}}

\def\gammabar{{\bar \gamma}}

\def\ibar{{\bar\imath}}
\def\jbar{{\bar\jmath}}
\def\kbar{{\bar k}}

\def\zbar{{\bar z}}
\def\Zbar{{\bar Z}}
\def\partialbar{{\bar\partial}}
\def\nablabar{{\overline{\nabla}}}

\newcommand{\newsection}[1]{\section{#1}\setcounter{equation}{0}}
\newcommand{\nn}{\nonumber}
\begin{document}

\rightline{DFPD-2012/TH/08}
\vskip 1cm
\centerline{\large \bf  Holomorphic Chern-Simons theory coupled to  off-shell Kodaira-Spencer gravity}

\vspace{1cm}

\vspace{1cm}

\centerline{    
  \textsc{ Stefano Giusto$^{1,2,a}$, ~Camillo Imbimbo$^{3,4,b}$, ~Dario Rosa$^{5,6,c}$}  }

\vspace{0.5cm}

\begin{center}
$^1\,$  Dipartimento di Fisica e Astronomia ``Galileo Galilei'',  Universit\`a di Padova,  Via Marzolo 8, 35131 Padova, Italy\\
$^2\,$ INFN, Sezione di Padova, Via Marzolo 8, 35131, Padova, Italy\\

\vspace{0.3 cm}

$^3\,$ Dipartimento di Fisica, Universit\`a di Genova,
Via Dodecaneso 33, 16146 Genoa, Italy\\
 $^4\,$ INFN, Sezione di Genova, Via Dodecaneso 33, 16146, Genova, Italy\\
\vspace{0.3 cm}

$^5\,$ Dipartimento di Fisica, Universit\`a di Milano-Bicocca, I-20126 Milano, Italy\\
$^6\,$ INFN, Sezione di Milano-Bicocca, I-20126 Milano, Italy
		    
\end{center}

\begin{center}
{\small $^a$stefano.giusto@pd.infn.it, ~~ $^b$camillo.imbimbo@ge.infn.it, ~~$^c$dario.rosa@mib.infn.it}
\end{center}

\vspace{1cm}

\centerline{ 
 \textsc{ Abstract}}

\vspace{0.2cm}

{\small
We construct an action for holomorphic Chern-Simons theory that couples the gauge field to off-shell
gravitational backgrounds, comprising the complex structure and the (3,0)-form of the target space.
Gauge invariance of the off-shell action is achieved by enlarging the field space to include an appropriate 
system of Lagrange multipliers, ghost and ghost-for-ghost fields.  Both the BRST transformations and the 
BV action are compactly and neatly written in terms of superfields which include fields, backgrounds and their antifields. We show that the anti-holomorphic target space  derivative
can be written as a BRST-commutator on a functional space containing the anti-fields of both the dynamical
fields and the gravitational backgrounds. We derive from this result a Ward identity that determines the 
anti-holomorphic dependence of physical correlators. 

}

\thispagestyle{empty}

\vfill
\eject

\setcounter{footnote}{0}
\newsection{Introduction}

Holomorphic Chern-Simons theory (HCS)  \cite{Witten:1992fb} was introduced by Witten as the target space field theory describing the dynamics of  a stack of $N$  5-branes of  topological string theory of the $B$ type living on a Calabi-Yau complex 3-fold
$X$.  The action of HCS 
\bea
\Gamma =  \int_X \Omega \wedge {\rm Tr} \bigl( \frac{1}{2}\, A\,\partialbar_Z\, A + \frac{1}{3}\, A^3\bigr)
\label{hcsactionone}
\eea
is  a 6-dimensional analogue of the topological 3-dimensional Chern-Simons action \cite{Schwarz:1978cn}.
The gauge field $A$,  encoding the open string degrees of freedom,  is  a one-form  with values in the Lie algebra of $SU(N)$ of type $(0,1)$ with respect to the chosen complex structure on $X$
\bea
A = dZ^\ibar \, A_\ibar (Z,\Zbar) = dZ^\ibar \, A^a_\ibar (Z,\Zbar)\, T^a .
\eea
In the formula above $T^a$ are the $SU(N)$ generators and ${\rm Tr}$ is the trace in its fundamental representation. $\partialbar_Z$ is the Dolbeault operator relative to  complex coordinates $(Z^i, Z^\ibar)$ compatible with the chosen complex structure
\bea
\partialbar_Z =dZ^\ibar\, \frac{\partial}{\partial Z^\ibar}\,.
\eea

The HCS action (\ref{hcsactionone}) depends therefore on {\it two} different classical geometrical data.  One of them is  the complex structure that one  picks on $X$. The other is $\Omega$,  the  globally defined  holomorphic (3,0)-form on $X$ 
\bea
\Omega= \Omega_{ijk}(Z,\Zbar)  \, dZ^i\wedge dZ^j\wedge dZ^k\,= \rho(Z,\Zbar)\, \epsilon_{ijk}\, dZ^i\wedge dZ^j\wedge dZ^k\,,
\eea
which, for Calabi-Yau three-folds, is unique up to a rescaling. $\Omega$ and the complex structure on $X$ are in correspondence with the {\it closed} moduli parametrizing the closed string vacuum in which the 5-branes live.  Since the (3,0)-form $\Omega$ depends
on the complex structure on $X$, the moduli space of closed strings is the total space of a complex line bundle whose base is the moduli space of complex structures on $X$ and whose fiber is the holomorphic (3,0)-form.

To exhibit explicitly the dependence of the theory on the complex structure 
of $X$ it is convenient to introduce the Beltrami parametrization of the differentials $dZ^i$
\bea
dZ^i = \Lambda^i_j \, \bigl( dz^j + \mu^j_\jbar\, dz^\jbar\bigr)\,,
\label{beltramidiff}
\eea
where $(z^i , z^\ibar)$ is a {\it fixed} system of complex coordinates.  The Beltrami differential
\bea
\mu \equiv \mu^i\,\frac{\partial}{\partial z^i} \equiv \mu^i_\jbar\, dz^\jbar\,\frac{\partial}{\partial z^i}
\eea
 is a (0,1)-form with values in the holomorphic tangent $T^{(1,0)}\,X$.  The action (\ref{hcsactionone}) rewrites in the system of coordinates  $(z^i , z^\ibar)$ as follows
\bea
&&\Gamma_0(\Omega, \mu) = \int_X \Omega\wedge \bigl(\frac{1}{2}A\,\nablabar\,A + \frac{1}{3}\,A^3\bigr)\,,
\label{hcsactiontwo}
\eea
where
\bea
\nablabar \equiv \partialbar - \mu^i\,\partial_i\,,\qquad \partialbar \equiv dz^i\, \frac{\partial}{\partial z^i}\,.
\eea
In this formulation,  the dependence of the theory on the closed moduli is captured by the two {\it classical} backgrounds fields ---  
$\Omega$ and $\mu$.   

The original action (\ref{hcsactionone}) is invariant under $\Omega$-preserving holomorphic reparametrizations.
The coupling of  $A$  to the classical background $\mu$ promotes this {\it global} invariance 
into a {\it local} symmetry under which $\mu$ transforms as a gauge field
\bea
&&s_{\text{diff}}\, \mu^i = -\partialbar \,\xi^i  + \xi^j\partial_j \,\mu^i - \partial_j\,\xi^i \,\mu^j \,.
\label{localdiffeos}
\eea
In (\ref{localdiffeos}), $\xi^i$ is the ghost of $\Omega$-preserving local diffeomorphisms
\bea
 \partial\, i_\xi(\Omega)=0\,,
\eea
where $i_\xi$ is the contraction of a form with the vector field $\xi^i\,\partial_i$. 

The  backgrounds $\Omega$ and $\mu$ in (\ref{hcsactiontwo}) must satisfy 
the classical equations of motion of the closed topological string theory:
\bea
&&  \mathcal{F}^i\equiv \partialbar \mu^i- \mu^j\partial_j\mu^i  =0\,,
\label{ksequation}\\
&&  \hat{\nablabar}\,\Omega \equiv \nablabar\Omega + \partial_i\, \mu^i\, \Omega =0\,.
\label{holo3form}
\eea
The first of such equations  is the celebrated Kodaira-Spencer equation \cite{Bershadsky:1993cx}  which expresses  the integrability of
the Beltrami differential; the second equation expresses the holomorphicity of  $\Omega$ in the complex structure
associated to $\mu^i$.   Indeed the action (\ref{hcsactiontwo})  is invariant under the  gauge  BRST symmetry\footnote{In this paper we will adopt the convention that fields and operators carrying odd ghost number {\it anti-commute} with fields and operators carrying odd form degree. In particular, the BRST operator $s$ and the Dolbeault differential $\nablabar$ anti-commute.}

\bea
&&s\, A =- \nablabar c - [A, c]_+\,,\nn\\
&& s\, c= -c^2\,,
\label{gaugetr}
\eea
where $c = c^a\, T^a$ is the anti-commuting ghost associated to $SU(N)$ gauge transformations, {\it only if}
the closed string equation of motions  (\ref{ksequation}) and (\ref{holo3form}) are satisfied.  It should be kept in mind
that $A$  and $c$ are  the dynamical variables of HCS while $\mu^i$, $\Omega$ and $\xi^i$ are classical non-dynamical fields.

For the purpose of investigating the {\it quantum} properties of HCS field theory, like its renormalization and its anomalies, it is useful to extend both gravitational backgrounds $\mu$ and $\Omega$ to be generic  {\it off-shell} functions. Hence in this article 
we  will write down the appropriate generalization of the action (\ref{hcsactiontwo})  valid also when $\mu$ and $\Omega$ do not satisfy their equations of motion (\ref{ksequation}) and (\ref{holo3form}).  
Nevertheless, as mentioned above,  the closed string fields will still be treated as non-dynamical backgrounds. In the context of string theory  our result could help understanding  the back-reaction  of the 5-branes on the closed string vacuum, since
the presence of branes modifies the equation of motions (\ref{ksequation}) and (\ref{holo3form}) and puts the backgrounds off-shell.

The standard method  to go ``off-shell''  is to introduce new fields acting as Lagrange multipliers whose equations of motions
are precisely the closed string equations  (\ref{ksequation}) and (\ref{holo3form}) and whose gauge transformation properties are such that the action is gauge invariant even for off-shell backgrounds. This strategy has been adopted by the authors
of \cite{Berkovits:2004jj},\footnote{
A different method to couple HCS to off-shell gravitational backgrounds has been put forward in 
\cite{Bonelli:2010cu}. Contrary to our approach, the (3,0)-form $\Omega$ is not treated in \cite{Bonelli:2010cu} as a background independent of the complex structure $\mu$ and, correspondingly, the  $\Omega$ equation (\ref{holo3form}) is still implicitly assumed, much like in the treatment of \cite{Berkovits:2004jj}. Moreover, the strategy employed to lift the Kodaira-Spencer constraint (\ref{ksequation}) entails the inclusion among the dynamical fields of the (1,0) component of the gauge field, together with a series of satellite fields, thus introducing a large gauge redundancy and making the dependence on the complex structure $\mu$ fairly implicit.
}
 who were able to solve, so-to-say, half of the problem:  they introduced a Lagrange multiplier whose equation of motion is the Kodaira-Spencer equation (\ref{ksequation}), but they did not reformulate the second equation (\ref{holo3form}) in the same way.  We achieve this task in the present article.
 
The reason why the authors of  \cite{Berkovits:2004jj}, whose main focus is the closed target space field theory, have restricted
$\Omega$ to be holomorphic,  
has to do with the different status that  equations (\ref{ksequation}) and (\ref{holo3form}) enjoy in the Kodaira-Spencer field theory:  Eqs. (\ref{ksequation}), which are the classical equations of motion derived from  the Kodaira-Spencer action \cite{Bershadsky:1993cx}, are equivalent to
the BRST-invariance of  the closed vertex operators associated to the complex structure moduli. This is the
standard relation between the second quantized classical equations of motion and first-quantized vertex operators. 

Eq.  (\ref{holo3form}), instead,  is not an equation of motion of Kodaira-Spencer field  theory. The $\Omega$ which enters the Kodaira-Spencer action must be holomorphic  and hence it is a  parameter and not a dynamical  field of Kodaira-Spencer theory.  From this
point of view, Kodaira-Spencer theory does not provide the second quantized formulation for the first-quantized vertex operator (of non-standard world-sheet ghost number)  associated to $\Omega$.

On the other hand, in the open string field theory it seems to be perfectly sensible to treat $\Omega$ and $\mu$ on the same footing: we will therefore 
introduce Lagrange multipliers whose equations of motion coincide with both  (\ref{ksequation}) and (\ref{holo3form})
and will determine their gauge transformation properties.  We will find it necessary to enlarge the $SU(N)$ gauge symmetry
to include  a number of  new ghost (and ghost-for-ghost) fields  which can be thought of as ``descendants''  of the Lagrange multipliers and which ensure the nilpotency of the full BRST transformations.

Since  $\Omega$ becomes, in our construction, an off-shell background, the HCS action that we will derive 
enjoys a larger reparametrization invariance than the original action (\ref{hcsactiontwo}).   This invariance include reparametrizations which are not $\Omega$ preserving:
\bea
&&s_{\text{diff}}\, \mu^i = -\partialbar \,\xi^i  + \xi^j\partial_j \,\mu^i - \partial_j\,\xi^i \,\mu^j\,, \nn\\
&&s_{\text{diff}}\, \Omega = \partial\, i_\xi(\Omega)\,,\nn\\
&& s_{\text{diff}} \, A = \xi^i\partial_i\, A\,,\qquad  s_{\text{diff}}\,\xi^i = \xi^j\,\partial_j\,\xi^i\,,
\label{chiraldiffeosone}
\eea
together with analogous transformations for all the other dynamical fields that we will introduce. We will refer to the
reparametrization invariance (\ref{chiraldiffeosone}) acting on off-shell $\mu^i$ and $\Omega$ as  {\it chiral} diffeomorphism
invariance. Chiral diffeomorphisms will be further discussed in Section \ref{section:reparametrizationinvariance}.

In Section \ref{section:gaugeinvariance} we write down the HCS gauge-invariant action coupled to off-shell gravitational backgrounds $\mu$ and $\Omega$ and the nilpotent BRST transformations acting on Lagrange multipliers
and ghost for ghosts.  

In Section \ref{section:superfields} we show that all fields and backgrounds of the theory, together with their anti-fields, belong
in superfields which are the sum of fields with different form and ghost degree and   have simple and compact BRST transformations rules. 

In Section \ref{section:theaction} we rewrite also the BV action of the theory in terms of superfields: we find that the full BV action
is obtained from the classical HCS action by promoting both fields and backgrounds to the superfield that each
of them belong to.
 
In the last Section of this paper, building on the superfield formulation of the theory,  we uncover an extended N=2 supersymmetric structure which underlies the off-shell HCS theory. We show that
the anti-holomorphic target space derivative $\partial_\ibar$ can be written as the (anti)-commutator of the gauge BRST operator 
with a supersymmetry charge $G_\ibar$ which acts on the space of all the dynamical fields and the gravitational backgrounds together with their anti-fields.  From this we derive a Ward identity which controls the anti-holomorphic dependence of physical correlators: the detailed analysis of the implications of this identity for the quantum properties of HCS is left to the future.

  \newsection{Chiral reparametrization invariance}
\label{section:reparametrizationinvariance}

The coupling of HCS to the holomorphic Beltrami differentials (\ref{beltramidiff}) is determined by requiring invariance
under  chiral reparametrizations. Chiral reparametrizations  act on the Beltrami differentials as follows 
\bea
&&s_{\text{diff}}\, \mu^i = -\partialbar \,\xi^i  + \xi^j\partial_j \,\mu^i - \partial_j\,\xi^i \,\mu^j \,,
\label{chiraldiffeos}
\eea
where $\xi^i$ is the anti-commuting ghost field of chiral diffeomorphisms:
\bea
&&s_{\text{diff}}\,\xi^i = \xi^j\,\partial_j\,\xi^i\,.
\eea
It is important to keep in mind that  $s_{\text{diff}}$ is nilpotent for {\it generic} $\mu^i$, independently of the validity of
the Kodaira-Spencer equation (\ref{ksequation}). On the space of  Beltrami differentials  $\mu^i$ which do satisfy Eq. (\ref{ksequation}) there exists a natural action of non-chiral (standard)  reparametrizations which follows from the definition (\ref{beltramidiff}): one can show \cite{Becchi:1987as}  that the actions of chiral and non-chiral reparametrizations coincide on such space if one identifies the
chiral ghost $\xi^i$ with the following combinations of the ghosts $(c^i, c^\ibar)$  of standard diffeomorphisms
\bea
\xi^i = c^i + \mu^i_\jbar\, c^\jbar \,.
\eea
There is no notion of standard reparametrizations of  ``off-shell''  Beltrami differentials, i.e. of $\mu^i$'s  which do not satisfy the
Kodaira-Spencer equation:  invariance under chiral diffeomorphisms (\ref{chiraldiffeos}) represents the  extention 
of reparametrization invariance appropriate  for off-shell $\mu^i$. 

Matter fields   with only anti-holomorphic indices  transform under  chiral diffeomorphisms as scalars
\bea
s_{\text{diff}}\, \phi_{\ibar\jbar\ldots} = \xi^i\,\partial_i \,\phi_{\ibar\jbar\ldots}\,.
\eea
For example,  the transformation law under chiral reparametrizations of the gauge field $A= A_\ibar\,dx^\ibar$ is 
\bea
s_{\text{diff}}\, A_\ibar = \xi^i\,\partial_i \, A_\ibar\,.
\eea
The action of chiral diffeomorphisms on tensors with holomorphic indices is instead
\bea
s_{\text{diff}}\, \phi^{i\,\ldots}_{\ibar\jbar\ldots; k\ldots} = \xi^j\,\partial_j \,\phi^{i\ldots}_{\ibar\jbar\ldots;k\ldots} - \partial_j\,\xi^i\, \phi^{j\ldots}_{\ibar\jbar\ldots;k\;\ldots}+\partial_k\,\xi^j\, \phi^{i\ldots}_{\ibar\jbar\ldots;j\;\ldots}+\cdots\,.
\eea
For example, in the following Section we will introduce  the Lagrange multiplier $C_i= C_{i\ibar}\,dx^\ibar$ which transforms under chiral reparametrizations as follows
\bea
s_{\text{diff}}\,  C_{i\ibar}= \xi^j\,\partial_j \, C_{i\ibar}+\partial_i\,\xi^j\, C_{j\ibar}\,.
\eea
For chiral reparametrizations there is a natural definition of {\it covariant  anti-holomorphic} derivative
\bea
\hat{\nabla}_\kbar\, \phi^{i\,\ldots}_{\ibar\jbar\ldots; k\ldots}= \nabla_\kbar\,\phi^{i\,\ldots}_{\ibar\jbar\ldots; k\ldots}+\partial_j\,\mu_\kbar^i\, \phi^{j\,\ldots}_{\ibar\jbar\ldots; k\ldots}-\partial_k\,\mu_\kbar^j\, \phi^{i\,\ldots}_{\ibar\jbar\ldots; j\ldots}+\cdots\,.
\label{holoconnection}
\eea
There is instead no natural notion of covariant {\it holomorphic} derivative. However
the holomorphic derivative of a tensor with no holomorphic indices  is a tensor with one holomorphic lower index.\footnote{ ``Natural'' in this context means that the connection in Eq.  (\ref{holoconnection}) depends
only on $\mu^i$ and not on the choice of a metric.} 

We will use the notation
\bea
\hat{\nablabar} \equiv dx^\kbar\, \hat{\nabla}_\kbar \equiv \nablabar + \hat\Gamma\,,
\eea
where the connection $ \hat\Gamma$ denotes the appropriate tensor product  of matrices with holomorphic indices 
\bea
(\hat{\Gamma})^i_j  = dx^\kbar\, \partial_j\mu_\kbar^i 
\eea
acting on holomorphic tensors in the usual way. For example
\bea
\hat{\nablabar}\, V_i \equiv \nablabar V_i - \partial_i\,\mu^j\, V_j\,.
\eea

\newsection{Gauge invariance}
\label{section:gaugeinvariance}
The variation of the  HCS action
\bea
&&\Gamma_0 = \frac{1}{2}\int_X \Omega\,\text{Tr}\bigl(A\,\nablabar\,A +\frac{2}{3}\, A^3\bigr)
\eea
 under the  BRST gauge transformations
\bea
&&s\, A = -\nablabar\, c- [A, c]_+\,,\nn\\
&& s\, c =-c^2
\label{gaugetrasfone}
\eea
is:
\bea
&&s\, \Gamma_0 = \frac{1}{2} \int_X \Omega\,\text{Tr}\bigl(  \nablabar\, c\,\nablabar\,A +  A\,\nablabar^2\, c\bigr)=\nn\\
&&\qquad =  \frac{1}{2}\int_X \Omega\,\nablabar\, \text{Tr} \bigl(c\,\nablabar\,A\bigr)+ \Omega\,\text{Tr}\bigl( c\, \nablabar^2\, A+  A\,\nablabar^2\, c\bigr)\nn\\
&&\qquad =  \frac{1}{2}\int_X \hat{\nablabar}(\Omega)\, \text{Tr} \bigl(c\,\nablabar\,A\bigr)+ \Omega\,\text{Tr}\bigl( c\, \nablabar^2\, A+  A\,\nablabar^2\, c\bigr)\,,
\label{gaugevariationone}
\eea
where
\bea
&&\hat{\nablabar}\,\Omega \equiv \nablabar\Omega + \partial_i\, \mu^i\, \Omega\,, \nn\\
&&\nablabar \equiv \partialbar  - \mu^i\,\partial_i \equiv  dx^\ibar\, \nabla_\ibar\equiv  dx^\ibar\,\bigl(\partial_\ibar  - \mu_\ibar^i\,\partial_i\bigr)\,.
\eea
The curvature of the $\nablabar$-differential  is
\bea
&&\nablabar^2  =  dx^\ibar\,dx^\jbar\,\bigl(\partial_\ibar  - \mu_\ibar^i\,\partial_i\bigr)\,\bigl(\partial_\jbar  - \mu_\jbar^j\,\partial_j\bigr)= -dx^\ibar\,dx^\jbar\,\bigl(\partial_\ibar\,\mu_\jbar^j-   \mu_\ibar^i\,\partial_i\,\mu^j_\jbar)\,\partial_j= -\mathcal{F}^i\,\partial_i \,,
\label{kscurvature}
\eea
where 
\bea
\mathcal{F}^j\equiv \partialbar\,\mu^j-   \mu^i\,\partial_i\,\mu^j
\eea
is the Kodaira-Spencer (0,2)-form with values in the holomorphic tangent.

Eq. (\ref{gaugevariationone})  shows that $\Gamma_0$ is gauge-invariant only if both $\Omega$ and $\mu^i$ are ``on-shell'', i.e. if they satisfy the  equations 
\bea
&&  \mathcal{F}^i\equiv \partialbar \mu^i- \mu^j\partial_j\mu^i  =0\,,\qquad \hat{\nablabar}\,\Omega \equiv \nablabar\Omega + \partial_i\, \mu^i\, \Omega =0\,.
\label{closedonshelleqs}
\eea
The first equation is equivalent to the nilpotency of $\nablabar$ while the second expresses the holomorphicity of $\Omega$ in the
complex structure defined by $\mu$. Let us introduce the Lagrange multipliers
\bea
C_i\equiv C_{i \ibar}\,dx^\ibar ,
\eea
a  (0,1)-form with values in the holomorphic cotangent, in corrispondence with the first of (\ref{closedonshelleqs}), and 
\bea
B \equiv  dx^\ibar\,\,dx^\jbar\, B_{\ibar\jbar} \,,
\eea
 a (0,2)-form, in correspondence with the second equation.

 If their BRST variations are taken to be
\bea
&&s\, B= -\text{Tr}\bigl( c\, \nablabar\, A\bigr)\,,\nn\\
&&s\, C_i=  \text{Tr}\bigl(-c\,\partial_i\, A +\partial_i\, c\,A\bigr)\,,
\label{gaugetrasftwo}
\eea
the action
\bea
\Gamma  = \frac{1}{2}\int_X \bigl[\Omega\,\text{Tr}\bigl(A\,\nablabar\,A+\frac{2}{3} A^3\bigr)+ \Omega\,\bigl(\nablabar\,B+ \mathcal{F}^i\,C_i \bigr)\bigr]
\label{hcsactionoff}
\eea
is  then BRST invariant for generic, ``off-shell'' backgrounds $\Omega$ and $\mu^i$.\footnote{The gauge transformation laws of $C_i$
in Eq. (\ref{gaugetrasftwo}) differ from those given in \cite{Berkovits:2004jj} but they are equivalent to them when $\Omega$ is on-shell.}

The  BRST transformations (\ref{gaugetrasfone}) and (\ref{gaugetrasftwo}) are not nilpotent when acting on the
Lagrange multipliers
\bea
&& s^2 \, B =\text{Tr}(c\,\nablabar^2 \,c)- \nablabar\,\text{Tr}\, (A\,c^2) = -\mathcal{F}^i\,\text{Tr} ( c\,\partial_i\,c)- \nablabar\,\text{Tr}\,( A\,c^2)\,,\nn\\
&& s^2 \, C_i  =\text{Tr}\,( -c\,\partial_i \,\nablabar\,c +\partial_i\, c \nablabar c )- \partial_i\,\text{Tr}\,( A\,c^2)= \nablabar\,\text{Tr}\,(c\,\partial_i\, c) + 
\text{Tr}\, c\, [\nablabar, \partial_i]\, c- \partial_i\,(\text{Tr}\, A\,c^2) =\nn\\
&&\qquad = \hat{\nablabar}\,\text{Tr}\, (c\,\partial_i\, c) - \partial_i\,\text{Tr}( A\,c^2)\,,
\label{brstsquaredone}
\eea
where we made use of the relation
\bea
\bigl[\partial_i,\,\nablabar_\ibar\bigr]= -\partial_i\,\mu_\ibar^j\,\partial_j\,.
\eea

The lack of nilpotency of (\ref{gaugetrasftwo}) is due to the existence of  new local symmetries of the action (\ref{hcsactionoff})   
\bea
B\to B^\prime = B + \mathcal{F}^i\, f_i+ \nablabar \, d\,,\qquad C_i \to C^\prime_i = C_i -\hat{\nablabar}\, f_i+ \partial_i\,d\,,
\label{extrabosonicsym}
\eea
with parameters $d \equiv d_\ibar\, dx^\ibar$ and  $f_i$ which are, respectively, a (0,1)-form and a section of the holomorphic cotangent.
The transformations (\ref{extrabosonicsym}) are symmetries of the action  (\ref{hcsactionoff}) since they leave invariant
the combination
\bea
&&\nablabar\,B+ \mathcal{F}^i\,C_i\to \nablabar\,B+ \mathcal{F}^i\,C_i  +\bigl(   \nablabar(\mathcal{F}^i\, f_i) -\mathcal{F}^i\,\hat{\nablabar}\, f_i\bigr)+\bigl(\nablabar^2\, d+ \mathcal{F}^i\,\partial_i\,d\bigr) = \nn\\
&&\qquad = \nablabar\,B+ \mathcal{F}^i\,C_i \,.
\eea
In the equation above  we made use of  (\ref{kscurvature}) and of the Bianchi identity for $\mathcal{F}^i$:
\bea
&&0= \epsilon^{\ibar\jbar\kbar}\,\bigl[\nablabar_\ibar,\bigl[\nablabar_\jbar, \nablabar_\kbar\bigr]\bigr] =
 -\epsilon^{\ibar\jbar\kbar}\,\nablabar_\ibar\,\bigl(\mathcal{F}_{\jbar\kbar}^i\,\partial_i\bigr)+
 \epsilon^{\ibar\jbar\kbar}\,\mathcal{F}_{\jbar\kbar}^i\,\partial_i\,\bigl(\nablabar_\ibar\bigr)=\nn\\
 &&\qquad =
  -\epsilon^{\ibar\jbar\kbar}\,\nablabar_\ibar\,\mathcal{F}_{\jbar\kbar}^i\,\partial_i+
 \epsilon^{\ibar\jbar\kbar}\,\mathcal{F}_{\jbar\kbar}^i\,\bigl[\partial_i,\,\nablabar_\ibar\bigr]=
 -\epsilon^{\ibar\jbar\kbar}\,\bigl[\nablabar_\ibar\,\mathcal{F}_{\jbar\kbar}^i+
 \mathcal{F}_{\jbar\kbar}^j\,\partial_j\,\mu_\ibar^i\bigr]\,\partial_i\,,
\eea
which can equivalently be written as
\bea
\hat{\nablabar} \,\mathcal{F}^i=0\,.
 \label{bianchiks}
 \eea 
Henceforth the BRST transformations 
\bea
&&s\, A = -\nablabar\, c - [A, c]_+\,,\nn\\
&&s\, c =- c^2\,,\nn\\
&&s\, B= - \text{Tr}\, (c\, \nablabar\, A)- \mathcal{F}^i\, f_i- \nablabar\, d\,,\nn\\
&&s\, C_i=  \text{Tr}\,(-c\,\partial_i\, A +\partial_i\, c\,A) -\hat{\nablabar}\, f_i+ \partial_i\, d\,,\nn\\
&& s\, d=  \text{Tr}\, (A\, c^2)\,,\nn\\
&& s\, f_i = - \text{Tr}\, (c\,\partial_i\, c)\,,
\label{gaugetrasfthree}
\eea
where $f_i$ and $d$ are {\it anti-commuting} fields with  ghost number $+1$, are  nilpotent when acting
on $A, c, B$ and $C_i$. The transformations (\ref{gaugetrasfthree}) are however {\it not} nilpotent when acting on $d$  and $f_i$:
\bea
&&s^2\, d = -\frac{1}{3}\, \nablabar\, \text{Tr}\,c^3\,,\nn\\
&&s^2\, f_i = - \frac{1}{3}\, \partial_i\, \text{Tr}\,c^3\,.
\label{brstsquaredtwo}
\eea
The reason why  the BRST rules (\ref{gaugetrasfthree}) are not nilpotent on $d$ and $f_i$ can be traced back to the fact that
 the replacements
\bea
&&d \to d^\prime = d + \nablabar\, e\,,\qquad  f_i\to f_i^\prime = f_i +\partial_i\, e
\eea
leave unchanged the transformations of $B$ and $C_i$ in (\ref{gaugetrasfthree}).  Therefore,  by introducing a  scalar  {\it commuting}   ghost-for-ghost field $e$   of ghost number +2, we obtain at last  the fully nilpotent 
BRST transformations of the action (\ref{hcsactionoff})
\bea
&&s\, A = -\nablabar\, c - [A, c]_+\,,\nn\\
&&s\, c =-c^2\,,\nn\\
&&s\, B= -\text{Tr}\,(c\, \nablabar\, A)- \mathcal{F}^i\, f_i- \nablabar \, d\,,\nn\\
&&s\, C_i=  \text{Tr}\,(-c\,\partial_i\, A +\partial_i\, c\,A) - \hat{\nablabar}\, f_i+ \partial_i\,d\,,\nn\\
&& s\,d =  \text{Tr}\,( A\, c^2)-\nablabar\, e\,,\nn\\
&& s\, f_i = - \text{Tr}\,(c\,\partial_i\, c)+\partial_i\, e\,,\nn\\
&&s\, e=  \frac{1}{3}\, \text{Tr}\,c^3\,.
\label{gaugetrasffive}
\eea
The structure of these BRST transformations  is possibly made more transparent by the remark  that the $c$-dependent terms in the BRST variations of  $B$, $d$ and $e$ are  precisely the forms which appear in the BRST descent equations that are generated by the holomorphic Chern-Simons (0,3)-form and the on-shell $\nablabar$:
\bea
s\, \Gamma^{(0,3)}  =- \nablabar \, \Gamma^{(0,2)}\,,\!\!\!\quad s\, \Gamma^{(0,2)}  = -\nablabar \, \Gamma^{(0,1)}\,,\!\!\!\quad s\, \Gamma^{(0,1)}  = -\nablabar \, \Gamma^{(0,0)}\,,\!\!\!\quad s\, \Gamma^{(0,0)}  = 0\!\quad \text{if}\; \;\nablabar^2=0\,,
\label{decentnablabar}
\eea
where 
\bea
&& \Gamma^{(0,3)} = \text{Tr}\bigl( A\,\nablabar\,A +\frac{2}{3}\, A^3\bigr)\,, \nn\\
&&\Gamma^{(0,2)}= \text{Tr}\,(c\, \nablabar\, A)\,,\nn\\
&&  \Gamma^{(0,1)}=   - \text{Tr}\,( A\, c^2)\,,\nn\\
&& \Gamma^{(0,0)} =-\frac{1}{3}\, \text{Tr}\,c^3\,.
\label{HCScocycle}
\eea
Therefore,  when $\mathcal{F}^i=0$, the cocycle 
\bea
&& \tilde{\Gamma}^{(0,3)} =  \Gamma^{(0,3)} + \nablabar\, B\,, \nn\\
&&\tilde{\Gamma}^{(0,2)}= \Gamma^{(0,2)} + s\, B +\nablabar\, d =0\,,\nn\\
&& \tilde{\Gamma}^{(0,1)}=   \Gamma^{(0,1)} + s\, d + \nablabar\, e =0\,, \nn\\
&& \tilde{\Gamma}^{(0,0)} =\Gamma^{(0,0)}+ s\, e =0
\eea
is a solution of the descent equations (\ref{decentnablabar})  which  is BRST equivalent to the Chern-Simons cocycle (\ref{HCScocycle}) and whose $(0,3)$ component  is precisely the form which appears in the off-shell action (\ref{hcsactionoff}). 

Summarizing, the (0,3)-form which appears in the off-shell Chern-Simons action is the representative of the solution of the cohomology problem (\ref{decentnablabar})  which is characterized by the vanishing of  the components of lower form-degree: its top-form component is, when $\nablabar^2=0$,   $s$-invariant ---  not just $s$-invariant modulo $\nablabar$.  The terms in (\ref{gaugetrasffive}) involving $f_i$ and $C_i$ are necessary to make $ \tilde{\Gamma}^{(0,3)} + \mathcal{F}^i\,C_i$ $s$-invariant even when $\nablabar^2\not= 0$.

The action  (\ref{hcsactionoff}) contains only covariant anti-holomorphic derivatives  and  therefore is manifestly invariant under chiral diffeomorphisms of both fields and backgrounds
\bea
&& s_{\text{diff}}\, A =\xi^i\,\partial_i \, A\,,\qquad s_{\text{diff}}\, c= \xi^i\,\partial_i\,c\,,\nn\\
&&s_{\text{diff}}\, B=  \xi^i\,\partial_i\, B\,,\qquad s_{\text{diff}}\,  C_i= \xi^j\,\partial_j \, C_i+\partial_i\,\xi^j\, C_j\,,\nn\\
&&s_{\text{diff}}\,d=\xi^i\partial_i\, d\,,\qquad s_{\text{diff}}\,f=\xi^i\partial_i\, f\,,\qquad s_{\text{diff}}\,e=\xi^i\partial_i\, e\,,\nn\\
&&s_{\text{diff}}\, \mu^i=-\hat{\nablabar}\,\xi^i\,,\qquad s_{\text{diff}}\,\xi^i = \xi^j\,\partial_j\,\xi^i\,,\qquad  s_{\text{diff}}\, \Omega = \partial\, i_\xi(\Omega)\,.
\label{diffbrstfinal}
\eea
Moreover the gauge BRST transformations (\ref{gaugetrasffive}) are also manifestly covariant, since they are expressed in terms
of anti-holomorphic covariant derivatives and holomorphic derivative of chiral reparametrizations scalars. Therefore the off-shell action (\ref{hcsactionoff}) is invariant under  the nilpotent total  BRST operator $s_{\text{tot}}$
\bea
s_{\text{tot}} \equiv s_{\text{diff}}+ s\,,
\label{brsttotal}
\eea
which encodes both the $SU(N)$ gauge symmetry and the   global  $\Omega$-preserving holomorphic reparametrization symmetry of the original action (\ref{hcsactionone}).

\newsection{Anti-fields and the chiral N=2 structure of the BRST transformations}
\label{section:superfields}
It is known \cite{Axelrod:1991vq} that the structure of the BRST symmetry of 3-dimensional (real) CS theory becomes considerably more transparent
when one considers,  together with the gauge connection $A$ and the ghost field $c$,  also their corresponding
anti-fields $A^*$ and $c^*$, which are, respectively, a 2-form of ghost number -1 and a 3-form of ghost number
-2.  All these fields can be collected in one single superfield, a polyform: 
\bea
\mathcal{A} = c + A+ A^* + c^* \,,
\label{Asuperfield}
\eea
whose total grassmannian degree $f=n_{\text{ghost}} +n_{\text{form}}$,  the sum of ghost number $n_{\text{ghost}}$ and anti-holomorphic form degree $n_{\text{form}}$, is $f=+1$. 
The BRST transformations of both fields and anti-fields of the 3-dimensional CS theory write 
nicely in terms of $\mathcal{A}$ as follows
\bea
(s+ d)\, \mathcal{A} + \mathcal{A}^2=0\,.
\eea

In this Section we will see that a similar strategy of collecting fields in polyforms of given grassmann parity $f$ also   elucidates  the geometrical content of the  BRST transformations of the HCS theory coupled to off-shell gravitational backgrounds.
 
 Let us  first write down the BRST transformations of the anti-fields of the dynamical fields.\footnote{For a condensed introduction to anti-fields and the Batalin-Vilkovisky (BV) formalism see  \cite{Fuster:2005eg}.}   The anti-field of a (0,1)-form
 $A= A_\ibar\, dx^\ibar$ is naturally  a (3,2)-form, $A^*$, whose BRST variation is
\bea
s\, A^* =  -\Omega\, \nablabar\, A-\frac{1}{2}\,\hat{\nablabar}\Omega\, A +\cdots\,,
\label{Aantifield}
 \eea
where the dots denote the contribution from fields other than $A$.  In order to obtain an anti-field which
sits in the same superfield (\ref{Asuperfield}) as $c$ and $A$, 
it is convenient to introduce the holomorphic density $\rho$
\bea
\Omega = \rho\, \epsilon_{ijk} \,dz^i\,dz^j\,dz^k
\eea
and to pull out a factor of $\rho$ from the definition of  the anti-field $A^*$:
\bea
A^{*}\to \rho\, A^{*}\qquad c^{*}\to \rho\,c^{*}\,.
\label{redefineone}
\eea
The redefined  $A^*$ becomes a  (0,2)-form and 
(\ref{Aantifield}) gets replaced by
\bea
 s\, A^* =  -\nablabar\, A-\frac{1}{2}\frac{\hat{\nablabar}\,\rho}{\rho}\, A +\cdots\,.
 \eea
We will use the notation
\bea
\frac{\hat{\nablabar}\,\rho}{\rho} =\frac{\nablabar\,\rho - \partial_i\mu^i\,\rho}{\rho} \equiv \hat{\nablabar} \log\rho\,.
\eea
Redefining both  $A^*$ and $c^*$ in this way, we obtain for their BRST transformations the expressions
\bea
&& s\, A^*=-\nablabar\,A- A^2 -[c, A^*]_++ 2\,C^{*\;i}\,
\partial_i\,c+\nn\\
&&- \frac{1}{2}\,(\nablabar\log\rho)\,A- B^*\,\nablabar\,c
- \bigl(\nablabar\,B^* +(\nablabar\log\rho)\,B^*\bigr)\,c+\nn\\
&&\qquad+ \bigl(\partial_i\,C^{*\;i}+( \partial_i\log\rho)\,C^{*\;i}\bigr)\,c +c^2\, d^*\,,\nn\\
&&s\, c^* =-[c, c^*]_+ -\nablabar\,A^*-[A, A^*]_++ 2\,C^{*\;i}\,\partial_i\,A+ 2\,f^{*\;i}\,\partial_i\,c+\nn\\
&&\qquad -(\nablabar\log\rho)\, A^* -
B^*\,\nablabar A +\bigl(\partial_i\,C^{*\;i} +(\partial_i\log\rho)C^{*\;i}\bigr)\,A +\nn\\
&&\qquad +\bigl(\partial_i\,f^{*\;i}+ (\partial_i\,\log\rho)\,f^{*\;i}\bigr)\,c+[A, c]_+\, d^*+ c^2\, e^*\,.
\label{brstantione}
\eea
The Lagrange multipliers $B,d, e$  sit in a single superfield with $f=2$. Therefore the corresponding
anti-fields $B^*,d^*, e^*$ are holomorphic densities with $f=0$. The multipliers $C_i$ and $f_i$ have $f=+1$,
and thus $C^{*\;i}$ and $f^{*\;i}$ are holomorphic densities with $f=+1$. We will find it convenient to redefine
$C^{*\;i}$ and $f^{*\;i}$ by pulling out a factor of  $\rho$, as we did with $A^*$ and $c^*$, 
\bea
C^{*\;i}\to \rho\, C^{*\;i}\,,\qquad f^{*\;i}\to \rho\,f^{*\;i}\,,
\label{redefinetwo}
\eea
so that the new anti-fields $C^{*\;i}$ and $f^{*\;i}$ are forms of anti-holomorphic degree 2 and 3 respectively.
We obtain therefore for the BRST transformation laws of the anti-fields of the Lagrange multipliers:
\bea
 &&s\, B^*= -\frac{1}{2}\,\hat{\nablabar}\rho\,,\nn\\
 && s\,d^*= \partial_i\,(\rho\, C^{*\;i})- \hat{\nablabar}\,B^*\,,\nn\\
&&s\, e^* = \partial_i\,( \rho \,f^{*\;i})-\nablabar\, d^*\,,\nn\\
&&s\,C^{*\;i} =- \frac{1}{2}\,\mathcal{F}^i \,,\nn\\
&&s\, f^{*\;i} =  - \hat{\nablabar}\,C^{*\;i} -\frac{B^*\, \mathcal{F}^i+(\hat{\nablabar}\,\rho)\,C^{*\;i}}{\rho}\,.
\label{brstantitwo}
\eea

The BRST transformation laws  (\ref{brstantitwo}) make clear that  the three anti-fields $B^*,d^*, e^*$
can be put together with the holomorphic density $\rho$ to form a ``complete''  BRST multiplet 
\bea
 \mathcal{B}^* \equiv  \rho + 2\, B^*+ 2\,d^*+2\,e^*
\eea
containing components  of all degrees $n_{\text{form}}=0,1,2, 3$ which transform as follows:
\bea
&&s\, \rho =0\,,\nn\\
&&s\, (2\, B^*)= -\hat{\nablabar}\rho\,,\nn\\
 && s\,(2\,d^*)= -\hat{\nablabar}\,(2\,B^*)+ \partial_i\,(2\,\rho\, C^{*\;i})\,,\nn\\
&&s\, (2\,e^*) = -\nablabar\, (2\,d^*)+\partial_i\,(2\, \rho \,f^{*\;i})\,.
\label{brstantithree}
\eea

To form a complete multiplet out of  $C^{*\;i}$ and $f^{*\;i}$ we need a (0,0)-form of ghost number 1 and a (0,1)-form of ghost number 0 with values in the holomorphic tangent:  The natural candidates are $\xi^i$, the chiral reparametrizations ghost,  and $\mu^i$, the Beltrami differentials. This motivates  considering the  total BRST operator 
\bea
s_{\text{tot}}= s_{\text{diff}}+ s \,,
\eea
 which encodes both  the chiral reparametrizations invariance and the gauge symmetry  of HCS theory. 
Indeed, one can check that  by defining
\bea
  \mathbb{M}^i \equiv  \xi^i +\mu^i +2\, C^{*i} + 2\,(f^{*i} - \frac{2}{\rho}\,B^*\, C^{*i})\equiv \xi^i +\mu^i +2\, C^{*i} + 2\,f^{*i}_n\,,
\label{superbeltrami}
\eea
the transformation rules for $C^{*\;i}$ and $f^{*\;i}$ in (\ref{brstantitwo}) assume the form
\bea
 s_{\text{tot}}\,  \mathbb{M}^i =-\bigl(\partialbar\,\mathbb{M}^i -\mathbb{M}^j\,\partial_j\, \mathbb{M}^i\bigr)\,.
\label{brstgeoone}
\eea
This equation also reproduces the correct BRST transformations for $\xi^i$ and $\mu^i$. 
From the same equation it also follows that the anti-holomorphic  derivative acting on super-fields 
$ \Phi^{i\,\ldots}_{\ibar\jbar\ldots; k\ldots}$
\bea
\hat{\nabla}_\kbar( \mathbb{M})\, \Phi^{i\,\ldots}_{\ibar\jbar\ldots; k\ldots}\equiv \partialbar_\kbar\,\Phi^{i\,\ldots}_{\ibar\jbar\ldots; k\ldots}- \mathbb{M}^j\,\partial_j\, \Phi^{i\,\ldots}_{\ibar\jbar\ldots; k\ldots}+\partial_j\,\mathbb{M}_\kbar^i\, \Phi^{j\,\ldots}_{\ibar\jbar\ldots; k\ldots}-\partial_k\,\mathbb{M}_\kbar^j\, \Phi^{i\,\ldots}_{\ibar\jbar\ldots; j\ldots}+\cdots
\label{holosuperconnection}
\eea
is covariant under the transformations (\ref{brstgeoone}).  Moreover the covariant differential
\bea
\hat{\nablabar}(\mathbb{M}) \equiv dx^\kbar\,\hat{\nabla}_\kbar( \mathbb{M})
\eea
satisfies 
\bea
\{s_{\text{tot}}, \hat{\nablabar}(\mathbb{M})\} + \hat{\nablabar}(\mathbb{M}) ^2=0\,.
\eea
This means that the operator
\bea
\delta \equiv s_{\text{tot}}+\hat{\nablabar}(\mathbb{M})
\eea
is nilpotent:
\bea
\delta ^2=0\,.
\eea
It is easily seen that the transformations (\ref{brstantithree}) rewrite in terms of this super-covariant anti-holomorphic derivative as
\bea
s_{\text{tot}}\, \mathcal{B}^* = - \hat{\nablabar}( \mathbb{M})\, \mathcal{B}^*\,.
\label{brstgeoBstar}
\eea

The introduction of the flat super-Beltrami $ \mathbb{M}^i$ allows one to recast the BRST transformations of
the gauge supermultiplet $c, A, A^*, c^*$ in a form which is analogous to the transformations  (\ref{Aantifield})
of the three-dimensional theory.  Defining the modified anti-fields 
\bea
A^*_n = A^* -\frac{B^*}{\rho}\, A-\frac{ d^*}{\rho}\,c\,,\quad c^*_n = c^*- 2\,\frac{B^*}{\rho}\, A^*_n-\frac{d^*}{\rho}\,A- \frac{e^*}{\rho}\,c
\eea
and the superfield
\bea
\mathcal{A} \equiv c+ A+ A^*_n + c^*_n\,,
\eea
the transformations of the gauge multiplet in   (\ref{gaugetrasffive}) and (\ref{brstantione}) write as 
\bea
s_{\text{tot}}\, \mathcal{A} =-\nablabar(\mathbb{M})\,\mathcal{A} - \mathcal{A}^2\,.
\label{brstgeoA}
\eea
Let us turn to the BRST transformations of the Lagrange multipliers. To form a complete BRST multiplet $\mathcal{B}$ out of  $B,d, e$ we need  to introduce the anti-field $\rho^*$, with ghost
number -1 and anti-holomorphic form degree 3,  corresponding to the background $\rho$.  

Let us comment on the significance of BRST transformations of the backgrounds and of their anti-fields.
Backgrounds (or coupling constants) can appear both in the classical action and
in the gauge-fixing term.  Backgrounds which appear only in the gauge-fixing term are of course {\it unphysical}. It is convenient 
in various contexts to extend the action of the BRST operator on the  unphysical backgrounds by introducing  corresponding fermionic super-partners  to form trivial BRST doublets (see \cite{Imbimbo:2009dy} and references therein). 
The BRST variation of physical backgrounds (or coupling constants) must instead be put to zero since varying a physical coupling constant is, by definition, not
a symmetry.  Indeed in HCS theory  the gauge BRST transformations of the (physical) backgrounds $\rho$ and $\mu^i$ vanish, as indicated in (\ref{brstantithree}) and (\ref{brstgeoBstar}).
However in the BV formalism it is natural to consider also  the anti-fields corresponding to physical backgrounds. 
Anti-fields of backgrounds  do not appear in the BV action since the BRST variation of the physical backgrounds vanish.
Their BRST variations are naturally defined in the BV formalism  by the derivatives of the BV action with
respect to the backgrounds. For HCS theory the BRST variations of the anti-fields of $\rho$ and $\mu^i$ can be defined to
be
\bea
&& s\,\rho^* = -\frac{\partial \Gamma_{BV}}{\partial \rho}=-\text{Tr}\,\bigl( \frac{1}{2}\,A\,\nablabar\,A+ \frac{1}{3}\,A^3\bigr)- \frac{1}{2}\,\nablabar\,B- \frac{1}{2}\, \mathcal{F}^i\, C_i \,,\nn\\
&& \frac{1}{\rho}\, s\, \mu^*_i =  -\frac{1}{\rho}\,\frac{\partial \Gamma_{BV}}{\partial \mu^i}= \frac{1}{2}\,\bigl(-\text{Tr}\,(A\,\partial_i\,A)+\partial_i\,B -\hat{\nablabar}\, C_i-(\hat{\nablabar}\log\rho)\, C_i\bigr) +\nn\\
&&\qquad -\text{Tr}\,(A^*\, \partial_i\,c)- \frac{B^*}{\rho}\, (\text{Tr}\,(c\,\partial_i\, A) - \partial_i\,d+\hat{\nablabar} f_i)-  \nablabar \bigl(\frac{B^*}{\rho} \bigr)\,f_i + \nn\\
&&\qquad 
-C^{* j}\,\partial_{[i}\, f_{j]} +\bigl(\partial_j\,C^{*\;j}+ (\partial_j\log\rho)\,C^{*\;j}\bigr)\, f_i +\frac{d^*}{\rho}\,\partial_i\, e\,,
\label{BRSTantibackgrounds}
\eea
where $\Gamma_{BV}$ is the BV action.\footnote{In Eq. (\ref{BRSTantibackgrounds}) we defined the functional derivative
of $\Gamma_{BV}$ with respect to $\rho$ by keeping constant the {\it true} anti-fields $A^*, c^*, C^{*i}$and $f^{*i}$,  and not
the redefined ones in (\ref{redefineone}),(\ref{redefinetwo}).}

 The content of the relation (\ref{BRSTantibackgrounds}) 
is that the variations of the action with respect to the physical backgrounds are BRST-closed: since the BRST transformations
do depend on the backgrounds this is not self-evident but it is ensured by the general BV formula.
In the enlarged field space which includes anti-fields of backgrounds such variations are BRST-trivial.

The superfield which collects together  $B,d, e$  and  $\rho^*$ and has nice BRST transformation laws  turns out to be 
\bea
&&\mathcal{B} = e + d + B_n + 2\,\rho^*_n\,,
\eea
where
\bea
&&B_n\equiv  B- 2\, C^{*i}\,f_i- \text{Tr}\, (A^*_n\,c)\,,\nn\\
&&2\,\rho^*_n \equiv   2\,\rho^* - 2\, C^{*i}\, C_i -  2\, f^{*i}_n\, f_i- \text{Tr} \,(A^*_n\,A+c^*_n\,c)\,.
\eea
One can check that the BRST transformation laws  for $B,d, e$   rewrite in terms of $ \mathcal{B}$ as follows
\bea
 s_{\text{tot}}\,  \mathcal{B} =-\nablabar(\mathbb{M})\, \mathcal{B} +\frac{1}{3}\,\text{Tr} \,\mathcal{A}^3\,.
\eea
The Lagrange multipliers $C_i$ and $f_i$ sit in a  superfield   which   contains also a 2-form of ghost number -1 and 
a 3-form of ghost number -2 with values in the holomorphic cotangent.  Looking at (\ref{superbeltrami}) one sees that 
these should be identified with the anti-fields $\mu_i^*$ and $\xi_i^*$ of the backgrounds $\mu^i$ and $\xi^i$.  
Since $\mathbb{M}^i$ is  valued
in the holomorphic tangent, $\mathbb{M}^*_i$ is naturally a holomorphic density. Choosing its  components to be
\bea
\mathbb{M}_i^*=  \rho\, f_i + (\rho\,C_i+ 2\, B^*\,f_i) + 2\,\mu_i^* +  2\,\xi^*_i\,,
\label{antisuperbeltrami}
\eea
its BRST transformation writes
\bea
 && s_{\text{tot}}\,  \mathbb{M}_i^* =-\hat{\nablabar}(\mathbb{M})\, \mathbb{M}_i^*+ \mathcal{B}^*\,\partial_i\, \mathcal{B}-
 \mathcal{B}^*\,\text{Tr}\, \mathcal{A}\,\partial_i\,\mathcal{A} \,.
\eea

The BRST transformations of all fields and backgrounds and their anti-fields  write in
a nice compact form in terms of the coboundary operator $\delta$:
\bea
 &&\delta\,  \mathbb{M}^i +\mathbb{M}^j\,\partial_j\, \mathbb{M}^i =0\,,\nn\\
&& \delta \mathcal{A} + \mathcal{A}^2=0\,,\nn\\
 && \delta\,  \mathcal{B} =\frac{1}{3}\,\text{Tr} \,\mathcal{A}^3\,,\nn\\
&& \delta\, \mathbb{M}_i^*=  \mathcal{B}^*\,\partial_i\, \mathcal{B}-  \mathcal{B}^*\,\text{Tr}\, \mathcal{A}\,\partial_i\,\mathcal{A}\,, \nn\\
&&\delta\, \mathcal{B}^* =0\,.
\label{brstgeofinal}
\eea

Let us comment on the geometrical interpretation of the BRST transformations (\ref{brstgeofinal}). The first of (\ref{brstgeofinal})
tells us that the super-Beltrami field $\mathbb{M}^i$ has flat Kodaira-Spencer curvature with respect to
the differential $\delta$.  The second equation expresses the flatness of the gauge super-connection $\mathcal{A}$. Since
$\mathcal{A}$ is flat, the Chern-Simons polyform
\bea
{\Gamma}_{CS} =  \text{Tr} \bigl(\mathcal{A}\, \delta \, \mathcal{A} + \frac{2}{3}\,\mathcal{A}^3\bigr)= - \frac{1}{3}\, \text{Tr}\,\mathcal{A}^3
\eea
is a $\delta$-cocycle.  The third equation in (\ref{brstgeofinal}) says that such cocycle is $\delta$-exact, being
the $\delta$-variation of $\mathcal{B}$. Taking the $\partial_i$ derivative of this equation one obtains
\bea
\delta\, \partial_i\, \mathcal{B}= \text{Tr}\, \partial_i\, \mathcal{A}\, \mathcal{A}^2= \delta\,\text{Tr}\,\mathcal{A}\,\partial_i\, \mathcal{A}\,.
\eea
This means that $\Omega_i \equiv \partial_i\, \mathcal{B} -\text{Tr}\,\mathcal{A}\,\partial_i\, \mathcal{A}$ is  a $\delta$-cocycle
\bea
\delta\,\bigl(\partial_i\, \mathcal{B} -\text{Tr}\,\mathcal{A}\,\partial_i\, \mathcal{A}\bigr)=\delta\, \Omega_i =0\,.
\eea
The fourth equation in (\ref{brstgeofinal})
\bea
 \mathcal{B}^*\,\Omega_i = \delta\, \mathbb{M}_i^*
\eea
implies therefore
\bea
 \delta\,\mathcal{B}^*\,\Omega_i=0\,.
\eea
This is consistent with the  fifth equation  in (\ref{brstgeofinal}) and implies that $\Omega_i$ is also $\delta$-trivial
\bea
\Omega_i  = \delta\, \mathcal{C}_i\,,\qquad  \mathbb{M}_i^*\equiv \mathcal{B}^*\, \mathcal{C}_i\,.
\eea
\newsection{The action}
\label{section:theaction}
Not only the BRST transformations but also the action rewrites in a neat form  in terms of superfields.  
The BV action corresponding to the gauge invariant action  (\ref{hcsactionoff}) is
\bea
&& 2\,\Gamma_{BV} =\rho\, \text{Tr}\,\bigl( \,A\,\nablabar\,A+ \frac{2}{3}\,A^3\bigr)+ \rho\,\nablabar\,B+  \rho\,\mathcal{F}^i\, C_i -2\,
\rho\, A^*\,s\,A -2\,\rho\, c^*\,s\,c + \nn\\
&& -2\,B^*\,s\, B - 2\,d^*\,s\,d - 2\,e^*\,s\,e - 2\,\rho\,C^{*i}\, s\, C_i-2\, \rho\, f^{*i}\,s\,f_i\,,
\label{gaugeBVaction}
\eea
where we chose to think of $\Gamma_{BV}$ as a $(0,3)$-form with values in the holomorphic densities rather than  a $(3,3)$-form
as the notation in Eq.  (\ref{hcsactionoff}) implies.

We have seen that when working with the superfields it is natural to promote the gauge BRST operator to the total
$s_{\text{tot}}$ which includes the chiral diffeomorphisms, by introducing the chiral reparametrization ghost $\xi^i$ which
should be thought of as a background, in the same way as $\rho$ and $\mu^i$. The corresponding BV action has extra terms
with respect to the gauge BV action (\ref{gaugeBVaction}) which are proportional to the background $\xi^i$. It is this extended
action which writes most simply in terms of superfields. Of course one can always recover the gauge action  (\ref{gaugeBVaction}) by putting $\xi^i$ to zero. 

A direct computation shows that (the extended)  $\Gamma_{BV}$ is the $(0,3)$-component of the following polyform with values in the holomorphic densities
\bea
&& 2\,\Gamma_{BV} =-\mathcal{B}^*\,s_{\text{tot}}\,\mathcal{B}-\mathbb{M}_i^*\, s_{\text{tot}}\,\mathbb{M}^i-\mathcal{B}^*\,\text{Tr}\, (\mathcal{A}\, s_{\text{tot}}\, \mathcal{A})=\nn\\
&&\qquad =\mathcal{B}^*\,\text{Tr}\,\bigl( \mathcal{A}\, \hat{\nablabar}(\mathbb{M})\, \mathcal{A}+ \frac{2}{3}\, \mathcal{A}^3\bigr)+ \mathcal{B}^*\,\hat{\nablabar}(\mathbb{M})\,\mathcal{B}+\mathbb{M}_i^*\, \bigl(\partialbar\,\mathbb{M}^i -\mathbb{M}^j\,\partial_j\, \mathbb{M}^i\bigr) \,.
\eea
We see therefore that, in much the same way as it happens for 3d CS theory\cite{Axelrod:1991vq}, the BV action is obtained from the classical action (\ref{hcsactionoff}) by replacing every field and background with the superfield to which it belongs
\bea
&& A \to \mathcal{A} \,,\qquad B\to \mathcal{B}\,,\qquad  \rho\, C_i \to \mathbb{M}^*_i\,,\nn\\
&&  \mu^i\to \mathbb{M}^i\,,\qquad \rho\to\mathcal{B}^* \,.
\eea
 
 \newsection{Anti-holomorphic dependence of physical correlators}
 \label{section:wardidentities}
The  stress-energy tensor of a topological quantum field theory is a BRST anti-commutator
\bea
T_{\mu\nu} = \{ s, G_{\mu\nu}\}\,,
\eea
where $G_{\mu\nu}$ is the  supercurrent. If both $T_{\mu\nu}$ and $G_{\mu\nu}$ are conserved  one obtains a
corresponding  relation for the charges
\bea
P_\mu = \{ s, G_\mu\}\,,
\label{topalgebra}
\eea
where $P_\mu$ is the generator of translations and $G_\mu$ is a vector supersymmetry.  Since $P_\mu$ is implemented on local
fields  by  space-time derivatives 
\bea
\partial_\mu = \{s, G_\mu\}\,,
\label{trivialderivative}
\eea
the relation (\ref{topalgebra}) proves that correlators of  local observables of topological field theories are space-time independent.

HCS theory is, in a sense, semi-topological: it does not depend on the full space-time metric but only on the Beltrami differential
$\mu^i$. Consequently we expect that a holomorphic version of the relation (\ref{trivialderivative}) holds for HCS:
\bea
\hat{\nabla}_\ibar = \{s, G_\ibar\}\,.
\label{nablabartrivial}
\eea
In this section we want to explore the validity of such a relation. We will find that a suitable $G_\ibar$ does indeed exist if we enlarge the
functional space upon which $G_\ibar$ acts to include the anti-fields of both  the dynamical fields and the backgrounds $\mu^i$ and $\Omega$.

It is convenient to introduce a field $\gamma^\ibar(\zbar)$, which depends only on the anti-holomorphic coordinates $z^\ibar$ and define the scalar operator 
\bea
G_{\gammabar}= \gamma^\ibar\, G_\ibar\,,
\eea
which carries ghost number -1.  It turns out that a suitable $G_\gammabar$ which satisfies (\ref{nablabartrivial}) is defined by the following simple action on the  superfields that we introduced
in Section \ref{section:superfields}
\bea
&& G_\gammabar\, \mathcal{A} = i_\gammabar (\mathcal{A})\,, \qquad G_\gammabar\, \mathcal{B} = i_\gammabar (\mathcal{B})\,,\qquad G_\gammabar\,\mathcal{B}^* = i_\gammabar (\mathcal{B}^*)\,,\nn\\
&& G_\gammabar\, \mathbb{M}^{i} = i_\gammabar (\mathbb{M}^{i})\,, \qquad  G_\gammabar\, \mathbb{M}^*_i = i_\gammabar ( \mathbb{M}^*_i)\,,
\eea
where $i_\gammabar$ is the contraction of a form with the antiholomorphic vector field $\gamma^\ibar\,\partial_\ibar$. $G_\gammabar$ so defined is easily seen to satisfy the relation 
\bea
\{ s_{\text{tot}}, G_\gammabar\} = \{ i_\gammabar, \partialbar\}\,,
\label{n2algebra}
\eea
where $s_{\text{tot}}$ is the BRST operator which include both gauge transformations and chiral diffeomorphisms:
\bea
s_{\text{tot}} = s_{\text{diff}} + s\,.
\eea
Note that the gauge BRST operator $s$ acts trivially on the gravitational backgrounds ($\mu^i$, $\rho$, $\xi^i$).
Let us show that  (\ref{n2algebra}) implies (\ref{nablabartrivial}) for the dynamical fields. Indeed, let $\Phi$ be a field which is neither
$\mu^i$  nor $\xi^i$. We have
\bea
&& G_\gammabar\, s_{\text{diff}}\, (\Phi)  = G_\gammabar( \mathcal{L}_\xi\, \Phi ) = \mathcal{L}_{i_\gammabar(\mu)}\, \Phi -\mathcal{L}_\xi\, G_\gammabar(  \Phi  )\,, \nn\\
&& s_{\text{diff}}\, G_\gammabar\, (\Phi)  = \mathcal{L}_\xi\, G_\gammabar(  \Phi  )\,, \nn\\
&& \{ s_{\text{diff}} , G_\gammabar\} = \mathcal{L}_{i_\gammabar(\mu)}\, \Phi\,,
\eea
where $\mathcal{L}_\xi$ denotes the action of chiral diffeomorphisms with parameter $\xi^i$. 
Hence
\bea
 \{ s, G_\gammabar\}\, \Phi = \{ i_\gammabar, \partialbar\}\, \Phi  - \{ s_{\text{diff}} , G_\gammabar\}\, \Phi =  \{ i_\gammabar, \hat{\nablabar}\}\, \Phi \,,
\eea
which is equivalent to  (\ref{nablabartrivial}).  Note that on the backgrounds, we have instead
\bea
 \{ s , G_\gammabar\}\, \xi^i =0\,,\qquad  \{ s , G_\gammabar\}\, \mu^i = i_\gammabar(\mathcal{F}^i)\,.
 \eea
When writing  down explicitly $G_\gammabar$ on the component fields one verifies that 
its  action on the sector which does {\it not} include the Lagrange multipliers  $B$ and $C_i$ 
does not involve the antifields of $\mu_i^*$ and $\rho^*$:
\bea
&& G_\gammabar \,c = i_\gammabar (A)\,, \nn \\
&& G_\gammabar A = i_\gammabar (A^\ast) - i_\gammabar \bigl(\frac{B^\ast}{\rho} A\bigr) - i_\gammabar \bigl(\frac{d^\ast}{\rho}\bigr)\, c \,, \nn\\
&&G_\gammabar A^* = i_\gammabar(c^*) -2\,i_\gammabar\bigl(\frac{B^*}{\rho}\bigr)\, A^* -\frac{B^*}{\rho}\, i_\gammabar(A^*)+\nn\\
&&\qquad  +i_\gammabar\bigl(\frac{B^*}{\rho}\bigr)\, \frac{B^*}{\rho}\,A + \frac{B^*}{\rho}\,i_\gammabar\bigl(\frac{d^*}{\rho}\bigr)\,c +\bigl(\frac{B^*}{\rho}\bigr)^2\,i_\gammabar(A) \,,\nn\\
&&G_\gammabar c^* = -2\, i_\gammabar\bigl(\frac{B^\ast}{\rho}\bigr)\,c^\ast - \frac{d^\ast}{\rho}\, i_\gammabar (A^\ast)
 - i_\gammabar\bigl(\frac{d^\ast}{\rho}\bigr)\, \frac{B^\ast}{\rho}\, A \,, \nn\\
&&G_\gammabar \,(\rho )  = 2\,i_\gammabar(B^*)\,,\nn\\
&&G_\gammabar \, B^* =  i_\gammabar(d^*) \,,\nn\\
&&G_\gammabar \, d^* =   i_\gammabar(e^*)\,,\nn\\
&&G_\gammabar \, e^*= 0\,,\nn\\
&& G_\gammabar\, \mu^i = 2\, i_\gammabar(C^{*i})\,,\nn\\
&& G_\gammabar\, C^{*i} = i_\gammabar(f^{*i}) - 2\, i_\gammabar\bigl(\frac{B^*}{\rho}\bigr) \, C^{*i} - 2\, \frac{B^*}{\rho}\, i_\gammabar(C^{*i}) \,,\nn\\
&&G_\gammabar\, f^{*i} =- 2\, i_\gammabar\bigl(\frac{B^*}{\rho}\bigr) \, f^{*i} - 2\, \frac{d^*}{\rho}\, i_\gammabar(C^{*i}) \,,\nn\\
&& G_\gammabar\, e = i_\gammabar(d)\,,\nn\\
&& G_\gammabar\,d = i_\gammabar(B)- 2\, i_\gammabar(C^{* i})\, f_i -\text{Tr}\,\bigl( i_\gammabar(A^*)\,c\bigr)+ i_\gammabar\bigl(\frac{B^*}{\rho}\,\text{Tr}\,( A\, c)\bigr)\,,\nn\\
&& G_\gammabar\, f_i = i_\gammabar(C_i)\,.
\label{Gcomponentone}
\eea
The action of $G_\gammabar$ on $B$ and $C_i$ involves instead 
the anti-fields $\mu_i^*$ and $\rho^*$  whose BRST transformations we introduced in (\ref{BRSTantibackgrounds}):
\bea
&& G_\gammabar\, B =2\, i_\gammabar(\rho^*)- 2\,i_\gammabar(C^{* i})\, C_i
- i_\gammabar\bigl(\frac{d^*}{\rho}\bigr)\,\text{Tr}\,( A\, c)  + \frac{B^*}{\rho}\,\text{Tr}\,(A\, i_\gammabar(A)) +\nn\\
&&\qquad  -\text{Tr}\,(i_\gammabar(A^*)\,A)
 \,,\nn\\
&& G_\gammabar\, C_i =  2 \,i_\gammabar\bigr(\frac{\mu^\ast_i}{\rho}) - 2\, i_\gammabar\bigl(\frac{B^\ast}{\rho}\, C_i\bigr) - 2\,i_\gammabar \bigl(\frac{d^\ast}{\rho}\bigr)\, f_i
\label{Gcomponenttwo}\,.
\eea

The existence of $G_\gammabar$ therefore reflects the semi-topological character of the theory.  Since the relation (\ref{trivialderivative}) valid for topological theories is replaced in  HCS by (\ref{nablabartrivial}),  the  correlators of physical local
observables $O(z, \zbar)$  
\bea
F(z, \zbar) = \langle O(z, \zbar)\, \cdots\rangle \quad \text{with}\quad s\, O(z, \zbar) =0\,,
\eea
where the dots denote insertions of physical observables at space-time points other than $(z,\zbar)$, satisfy the identity
\bea
\hat{\nabla}_\ibar\, F(z, \zbar)  =  \langle s\, \bigl(G_\ibar\,O(z, \zbar)\bigr)\cdots \rangle\,.
\label{holowi}
\eea 
One cannot immediately conclude, from this Ward identity (and BRST invariance) that  $F(z, \zbar)$ is  a holomorphic function (tensor) on $X$.
This  for two reasons. 

First of all we have seen that $G_\ibar$ when acting on $B$ and $C_i$ produces
the $\mu^*_i$ and $\rho^*$: since  $\rho$ and $\mu^i$ are not dynamical (one does not integrate over them) the Ward identity
(\ref{holowi}) says that the $\zbar$-dependence of physical correlators involving $B$ and $C_i$ can be expressed in terms
of derivative of correlators with respect to the moduli  $\rho$ and $\mu^i$.

Secondly, even if restricted to observables which do not involve the Lagrange multipliers $B$ and $C_i$, the Ward identity (\ref{holowi}) ``almost'' implies the holomorphicity of $F(z, \zbar)$, but not quite. Indeed, the  $G_\gammabar$ variations  (\ref{Gcomponentone}) of  fields other than $B$ and $C_i$ contain the dynamical anti-fields, and the functional averages of  the BRST variation
of operators which depend on the anti-fields are, in general,  zero only up to contact terms.

At any rate it is clear that the Ward identity (\ref{holowi}) strongly constrains the anti-holomorphic dependence of physical
correlators. This equation should  therefore play for the Green functions of physical observables of  HCS field theory 
the role that the holomorphic anomaly equation plays for the open-closed topological string amplitudes \cite{Walcher:2007tp}.
For example, it is conceivable that one could determine, to a large extent,  the space-time dependence of physical correlators of HCS using the identity (\ref{holowi}) together with assumptions about the behavior of correlators at infinity. An analogous approach to  compute topological open and closed string amplitudes by integrating the
holomorphic anomaly equation has been quite successful \cite{Bershadsky:1993cx},\cite{Hosono:1999qc}.

The study of the full implications for the {\it quantum} properties of HCS field theory is left to the future. 
Here we will limit ourselves to few brief comments. First of all there is the issue of anomalies: the  chiral diffeomorphism symmetry (\ref{diffbrstfinal})  can, in principle, suffer from anomalies, and, indeed, it does \cite{Losev:1996up}.  Chiral diffeomorphism invariance can be restored at the price of introducing a dependence on the anti-holomorphic Beltrami differentials and, possibly, on the K\"ahler metric. The chiral diffeomorphism invariant theory should display an anomalous Ward identity which controls the anti-holomorphic dependence on the backgrounds very much like (\ref{holowi}) does for the space-time anti-holomorphic dependence.


But, of course, the real question which remains to be addressed is the ultraviolet completeness of the HCS quantum field
theory. Being a 6-dimensional gauge theory, HCS theory is superficially not renormalizable. On the other hand its string interpretation suggests the opposite. We believe that the extended supersymmetry  structure (\ref{nablabartrivial}) capturing the semi-topological character of the theory and the identity (\ref{holowi}) restricting the space-time dependence of quantum correlators  should be instrumental in ensuring that  the physical sector of the theory is indeed free of  ultra-violet divergences.

\section*{Acknowledgments.}

C.I.  would like to thank R. Stora for useful conversations and the theory group at CERN  for hospitality during various stages of this work.  He also thanks  the Isaac Newton Institute for
Mathematical Sciences of Cambridge, UK,  for hospitality and financial support during  the ``Mathematics and Applications of Branes in String and M-theory'' program.  This work is supported in part 
by the MIUR-PRIN contract 2009-KHZKRX, by the Padova University Project CPDA119349 and by INFN.

\providecommand{\href}[2]{#2}

\end{document}